\documentclass[11pt,a4paper,english,nofootinbib]{revtex4-2}
\usepackage{lmodern}
\usepackage{lmodern}

\usepackage[T1]{fontenc}
\usepackage[latin9]{inputenc}
\usepackage{babel}
\usepackage{calc}
\usepackage{amsmath}
\usepackage{amssymb}
\usepackage{graphicx}
\usepackage{esint}
\usepackage[unicode=true,pdfusetitle,
 bookmarks=true,bookmarksnumbered=false,bookmarksopen=false,
 breaklinks=false,pdfborder={0 0 1},backref=false,colorlinks=false]
 {hyperref}
\usepackage{xcolor}
\makeatletter

\@ifundefined{textcolor}{}
{%
 \definecolor{BLACK}{gray}{0}
 \definecolor{WHITE}{gray}{1}
 \definecolor{RED}{rgb}{1,0,0}
 \definecolor{GREEN}{rgb}{0,1,0}
 \definecolor{BLUE}{rgb}{0,0,1}
 \definecolor{CYAN}{cmyk}{1,0,0,0}
 \definecolor{MAGENTA}{cmyk}{0,1,0,0}
 \definecolor{YELLOW}{cmyk}{0,0,1,0}
}

\usepackage{babel}

\usepackage{graphicx}
\def\b{\begin{equation}}
\def\e{\end{equation}}

\@ifundefined{textcolor}{}{%
 \definecolor{BLACK}{gray}{0}
 \definecolor{WHITE}{gray}{1}
 \definecolor{RED}{rgb}{1,0,0}
 \definecolor{GREEN}{rgb}{0,1,0}
 \definecolor{BLUE}{rgb}{0,0,1}
 \definecolor{CYAN}{cmyk}{1,0,0,0}
 \definecolor{MAGENTA}{cmyk}{0,1,0,0}
 \definecolor{YELLOW}{cmyk}{0,0,1,0}
 }

\usepackage{latexsym}\usepackage{bm}

\begin{document}
\title{Perturbative solution of the Einstein Constraints with Spin and Momentum Far Away From a Binary Source in the Bowen-York formalism}
\author{{\normalsize{}{}{}{}{}{}{}{}{}{}{}{}{}{}{}{}{}Emel
Altas}}
\email{emelaltas@kmu.edu.tr}

\affiliation{Department of Physics, Karamanoglu Mehmetbey University, 70100, Karaman,
Turkey}

 \author{{\normalsize{}{}{}{}{}{}{}{}{}{}{}{}{}{}{}Emine Ertugrul}}
\email{emine.ertugrul@boun.edu.tr}

\affiliation{Department of Physics, Bogazici University,
34342 Bebek, Istanbul, Turkey }

\author{{\normalsize{}{}{}{}{}{}{}{}{}{}{}{}{}{}{}{}{}Bayram
Tekin}}
\email{btekin@metu.edu.tr}

\affiliation{Department of Physics, Middle East Technical University, 06800, Ankara,
Turkey}
\date{{\normalsize{}{}{}{}{}{}{}{}{}{}{{}{}{}{}\today}}}

\begin{abstract}

We study the momentum and Hamiltonian constraints of vacuum Einstein equations, within the Bowen-York formalism, for two interacting black holes in close separation,  with anti-parallel spins and anti-parallel linear momenta. We give an analytical solution using perturbation theory. We also compute the location and the shape of the apparent horizon which generically depend on all the parameters, angles and the separation between the black holes.  Our solution only works for distances far away from the black holes. To gain more insight close to the black holes, one has to go to the higher orders in perturbation theory, which is a rather cumbersome process. But the solution presented here  can be of some use for numerical computations as the latter should match our result for the described problem.
\end{abstract}

\maketitle

\section{Introduction}

The days are long gone when black holes are merely a theoretical curiosity that necessarily show up in General Relativity but are thought, somehow, would not appear as final states of gravitational collapse.  They are now in the domain of observation either as sources of gravitational waves \cite{merger,intermediate} due to their merger, about one hundred of such mergers have been detected; or as sources of strong gravitational lensing which yields a characteristic shadow in the observer's screen granted that black hole has a non-vacuum environment such as its accretion disk or photon sphere. This is the case for the supermassive black holes at the center of  galaxies, two of which have been directly observed \cite{EHT1, EHT2}. The second one being the image of the supermassive black hole in our own galaxy.

In the LIGO/VIRGO type detectors, two observables are recorded: these are  
the time-dependent gravitational wave frequency $\nu(t)$, and the strain caused by the wave $h(t)$.  From these two data, using General Relativity, one can derive all the information, except its exact location in the sky, regarding the source of the transient gravitational wave. So to be able to interpret the observed data, General relativity is still, in some sense, our greatest tool. So far, the theory is consistent with all the observations. Incidentally, with a slight modification, Newtonian physics \cite{Mathur, Thorne} also helps us understand the basic aspects of the black hole merger physics and reproduce the properties, such as the masses of individual black holes and the power radiated during the merger \cite{Oz}. To be able to understand the profile of the gravitational wave generated during the inspiral, merger and the ring-down phases, one necessarily  resorts to numerical relativity, especially in the merger phase for the largest strain and the largest wave frequency is generated. This was finally achieved in \cite{Pre} and for a nice brief account of this, see \cite{Shap}. Here we have nothing more to add to this well-tested, highly successful  numerical evolution scheme of Einstein's evolution equations. Instead we shall concentrate on not the evolution equations, but the constraint equations; and try to understand the binary black hole in a close orbit as an initial value problem. We shall discuss the importance of the Einstein constraint equations in the next section, but let us note that the constraints not only determine the possible initial data, but they also determine the time evolution of the system. But the constraints are extremely difficult to solve.

Recently \cite{Altas_Tekin_single}, we studied the constraint equations of General Relativity in the Bowen-York \cite{BY,Altas_redux} formalism and constructed approximate initial data (for the vacuum case) for a single black hole with spin and linear momentum pointing in arbitrary directions. Bowen-York approach in solving the Einstein constraint equations starts with a conformally flat 3-metric and hence the single gravitating object described in this formalism typically has junk gravitational radiation; and before it settles to a single black hole, it will emit this radiation.  Furthermore, if it is spinning, in the final stationary state, it cannot be represented in the conformal Bowen-York form \cite{Garat, Kroon}.  It would be pedantic to stress the importance of understanding the merger of black holes as we are living in a time, observation of not only black hole collisions but also other compact objects is in a thrilling state since the first announcement \cite{merger}.

In this work, we extend our earlier discussion to binary black holes, still in the Bowen-York formalism, orbiting around each other with generically different but anti-parallel spins and linear momenta. We assume that the spacetime is asymptotically flat and globally hyperbolic with conformally flat hypersurfaces as in \cite{BY}, which makes the momentum constraint easily solvable. But of course the Hamiltonian constraint is a nonlinear partial differential equation which ultimately requires numerical techniques to be solved. Here, instead, we use perturbation theory, assuming small spin and small linear momenta and separation (compared to masses and the distances we are looking at) to obtain analytical formulas for the conformal factor of the spatial metric. We also compute the shape and location of the apparent horizon. Our perturbative approach fails for distances close to the black holes.

The layout of the paper is as follows: in section II we describe the constraint equations and the initial data for two black holes by solving the momentum constraint; and find the form of the Hamiltonian constraint that defines our system.
That section should be considered as background material and the only original expression is given as (\ref{BYkare}).
 Fig. (\ref{fig:Positions0}) summarizes our assumptions. We also compute the ADM momentum and spin from the extrinsic curvature. For these two quantities, only the asymptotic form of the conformal factor is needed, and hence the computation can be carried out before solving the Hamiltonian constraint, with the knowledge of the extrinsic curvature.
In section III, we find the approximate solution to the Hamiltonian constraint and in Section IV we compute the shape of the apparent horizon.  We also compute the irreducible mass, the area of the apparent horizon and the ADM mass of the system.    

\section{The Einstein Constraint equations and the initial data}

Let us consider a spatial 3-dimensional hypersurface  $\Sigma$ embedded in a globally hyperbolic, asymptotically flat spacetime; then from Einstein's equations in a vacuum, one obtains the following constraints (see for example \cite{Bartnik})  
\begin{eqnarray}
 &&-^{\Sigma}R-K^{2}+K_{ij}K^{ij}=0,\nonumber \\
 && 2D_{k}K_{i}^{k}-2D_{i}K=0,~~~~~~~~~~~~~~\label{Einstein_c}
\end{eqnarray}
where $K_{i j}$ is the extrinsic curvature of the hypersurface and $^{\Sigma}R$ is the scalar curvature constructed from $\gamma_{ij}$, the metric on  $\Sigma$; and the trace of the extrinsic curvature is $K= \gamma^{i j} K_{i j}$.  Here $D_{i}$ is the covariant derivative compatible with $\gamma_{i j}$.

These constraint equations, together with the first order time evolution equations,    which we do not depict here explicitly, constitute a dynamical system formulation of Einstein's equations. What is remarkably beautiful is that the linearization of the constraints (\ref{Einstein_c}) appear in the time-evolution equations as was given by Fischer and Marsden \cite{FM}
\begin{equation}
\frac{d}{dt}\begin{pmatrix}\gamma\\
\pi
\end{pmatrix}=J\circ D\Phi^{*}(\gamma,\pi)({\cal {N}}),\label{evolution}
\end{equation}
where the $J$ matrix reads 
\begin{equation}
J=\begin{pmatrix}0 & 1\\
-1 & 0
\end{pmatrix}.
\end{equation}
Here the canonical momentum, a tensor density, $\pi$ is linearly related to the extrinsic curvature as $\pi^{ i j}= \sqrt{\gamma}( K^{i j}- \gamma^{i j}K)$; and $D\Phi^{*}$ is the formal adjoint of the linearized constraints and ${\cal {N}}$ is the lapse-shift four vector. [Note that, this $D$ operator is not to be confused with the covariant derivative $D_i$.]
We invite the interested reader to follow a detailed derivation of these equations from scratch in 
\cite{our_dain_paper}. The up-shot here is, as mentioned in the introduction part, that the constraints play a dual role: they determine the initial data and the time evolution, hence they are extremely important in General Relativity. This point of view was stressed in \cite{Bartnik}. But they are extremely hard to solve. 

In seeking for solutions of the constraint equations, there are various ways to adopt, some of which are well-described in \cite{Eric0}. We shall follow the Bowen-York approach \cite{BY}  and assume "maximal slicing" ($K=0$), and a  conformally flat hypersurface: $\gamma_{ij}=\psi^{4}f_{ij}$
with $f$ being the flat metric in some coordinates on the hypersurface. Then the Einstein constraints (\ref{Einstein_c}) reduce to 
\begin{eqnarray}
 &  & \hat{D}_{i}\hat{D}^{i}\psi=-\frac{1}{8}\psi^{-7}\hat{K}_{ij}^{2},\label{elliptic0}\\
 &  & \hat{D}^{i}\hat{K}_{ij}=0,\label{trans}
\end{eqnarray}
with $\hat{D}_{i}f_{jk}=0$ and $K_{ij}=\psi^{-2}\hat{K}_{ij}$. The momentum constraint (\ref{trans}) nicely decouples and is amenable to exact analytical solution. Note that here we are just summarizing what is already known in this system, we do not claim to suggest a novel approach.

\begin{figure}
\centering \includegraphics[width=0.8\linewidth]{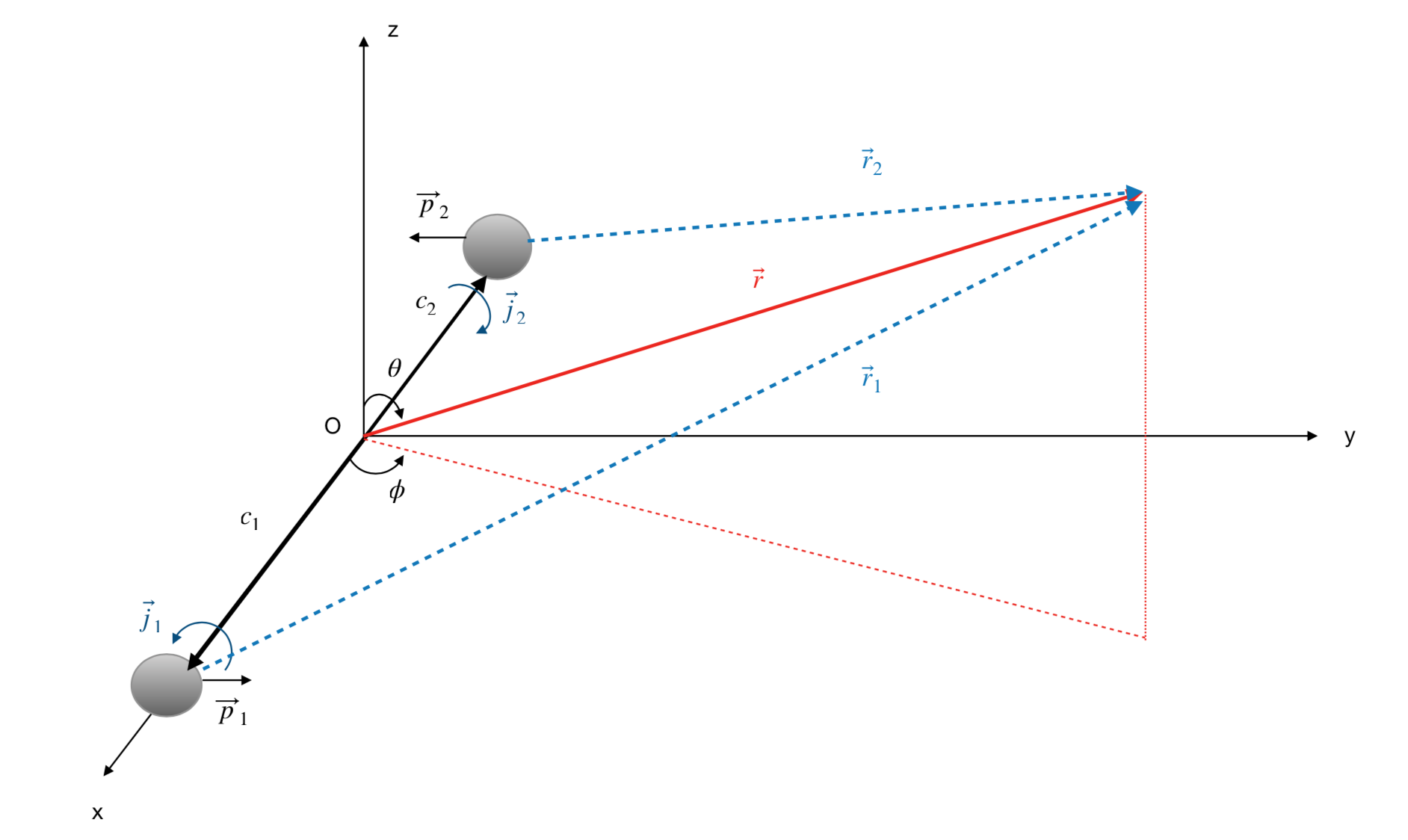}\caption{Positions of 
two gravitating black holes: $p_i$ are the linear momenta pointing in the $\pm \hat{y} $ direction, $j_i$ are the spins in the $\pm \hat{x} $ direction; while $c_i$ show the locations of the black holes from the center of the coordinates. The linear momenta are anti-parallel  to each other and the spins are anti-parallel to each other and lie along the axis connecting the black holes. The linear momenta and spins are perpendicular to each other. This configuration is chosen to mimic the black hole merger event.}
\label{fig:Positions0} 
\end{figure}

There could be many possible solutions to (\ref{trans}): following Bowen-York (\cite{BY}, we choose the following solution which can be interpreted (as later justified from the total ADM linear and angular momentum computations) as two gravitating objects located at different points in a vacuum as depicted in Fig. (\ref{fig:Positions0}). Note that we shall take the linear momenta to be anti-parallel to describe the circular motion about the center of mass; and the spins to be anti-parallel and perpendicular to linear momenta as within General Relativity spin-spin interactions lead to such a configuration.  At the perturbative level in General Relativity, the relevance of this configuration can be seen easily \cite{Gullu,Tasseten}.
Hence we take the following form for the scaled extrinsic 
curvature
\begin{eqnarray}
 &&\hat{K}_{ij}=\frac{3}{2r_{1}^{2}}\Big(p_{1i}n_{1j}+p_{1j}n_{1i}+(n_{1i}n_{1j}-f_{ij})\,p_{1}\cdot n_{1}\Big) +\frac{2}{r_{1}^3} \Big( \left( j_{1}\times n_{1}\right) _{i} n_{1j}+\left(  j_{1}\times n_{1}\right) _{j} n_{1i}\Big) \nonumber\\
&+& \frac{3}{2r_{2}^{2}}\Big(p_{2i}n_{2j}+p_{2j}n_{2i}+(n_{2i}n_{2j}-f_{ij})\,p_{2}\cdot n_{2}\Big) +\frac{2}{r_{2}^3} \Big( \left( j_{2}\times n_{2}\right) _{i} n_{2j}+\left( j_{2}\times n_{2}\right)_{j} n_{2i}\Big), \nonumber\\
 \label{Bowen-York_extrinsiccurvature}
\end{eqnarray}
where $r_{1}, r_{2}>0$ are distances from the centers of the black holes,  $n_{1i}$ and  $n_{2i}$  are unit normals
on spheres of radii $r_{1}, r_{2}>0$ . [For a more  general Bowen-York type solution than the one we have taken here, see \cite{Beig}.]
To solve the  Hamiltonian constraint equation (\ref{elliptic0}), we need to find the square of the extrinsic curvature (\ref{Bowen-York_extrinsiccurvature}). Here we shall use the vector notation as it, otherwise, gets cumbersome. The below formula is valid for {\it generic} angles, not just for the particular case depicted in  Fig. (\ref{fig:Positions0}).
\begin{eqnarray}
&&\hat{K}_{ij}\hat{K}^{ij}= 
\frac{9}{r_{1}^{4}}\left( \vec{p}_{1}^{2}+2(\vec{p}_{1}\cdot \vec{n}_{1})^{2}\right)+\frac{9}{r_{2}^{4}}\left( \vec{p}_{2}^{2}+2(\vec{p}_{2}\cdot \vec{n}_{2})^{2}\right)\nonumber\\
&+&\frac{9}{r_{1}^{2}r_{2}^{2}}\big(\vec{n}_{1}\cdot \vec{n}_{2}\left[  \vec{p}_{1}\cdot \vec{p}_{2}  + \left(\vec{p}_{1}\cdot \vec{n}_{2}\right)\left(\vec{p}_{2}\cdot \vec{n}_{2}\right) +\left(\vec{p}_{1}\cdot \vec{n}_{1}\right) \left(\vec{p}_{2}\cdot \vec{n}_{1}\right) + \frac{1}{2} \left(\vec{p}_{1}\cdot \vec{n}_{1}\right)\left(\vec{p}_{2}\cdot \vec{n}_{2}\right) \left(\vec{n}_{1}\cdot \vec{n}_{2}\right)\right]  \nonumber\\
&+& \left(\vec{p}_{1}\cdot \vec{n}_{2}\right) \left(\vec{p}_{2}\cdot \vec{n}_{1}\right)-\frac{3}{2} \left(\vec{p}_{1}\cdot \vec{n}_{1}\right) \left(\vec{p}_{2}\cdot \vec{n}_{2}\right)\big)\nonumber\\
&+&\frac{18}{r_{1}^{5}} \left(\vec{j}_{1}\times \vec{n}_{1} \right)\cdot \vec{p}_{1}   +\frac{18}{r_{2}^{5}}\left( \vec{j}_{2}\times\vec{n}_{2}\right)  \cdot \vec{p}_{2}\nonumber\\
&+& \frac{18}{r_{1}^{3}\, r_{2}^{2}} \left( \vec{n}_{1}\cdot \vec{n}_{2} \left[ (\vec{j}_{1}\times \vec{n}_{1})\cdot \vec{p}_{2}+(\vec{j}_{1}\times \vec{n}_{1})\cdot \vec{n}_{2}(\vec{p}_{2}\cdot \vec{n}_{2})\right] + (\vec{j}_{1}\times \vec{n}_{1})\cdot \vec{n}_{2}(\vec{p}_{2}\cdot \vec{n}_{1}) \right)  \nonumber\\
&+& \frac{18}{r_{1}^{2}\, r_{2}^{3}}\left( \vec{n}_{1}\cdot \vec{n}_{2} \left[ (\vec{j}_{2}\times \vec{n}_{2})\cdot \vec{p}_{1}+(\vec{j}_{2}\times \vec{n}_{2})\cdot \vec{n}_{1}(\vec{p}_{1}\cdot \vec{n}_{1})\right] + (\vec{j}_{2}\times \vec{n}_{2})\cdot \vec{n}_{1}(\vec{p}_{1}\cdot \vec{n}_{2}) \right) \nonumber\\
&+&\frac{18}{r_{1}^{6}}(\vec{j}_{1}\times \vec{n}_{1})\cdot(\vec{j}_{1}\times \vec{n}_{1})+\frac{18}{r_{2}^{6}}(\vec{j}_{2}\times \vec{n}_{2})\cdot(\vec{j}_{2}\times \vec{n}_{2})\nonumber\\
&+&\frac{36}{r_{1}^{3}\, r_{2}^{3}}\left[ (\vec{j}_{1}\times \vec{n}_{1})\cdot(\vec{j}_{2}\times \vec{n}_{2})(\vec{n}_{1}\cdot \vec{n}_{2})+ (\vec{j}_{1}\times \vec{n}_{1})\cdot \vec{n}_{2} (\vec{j}_{2}\times \vec{n}_{2})\cdot \vec{n}_{1}\right].
\label{BYkare}
\end{eqnarray}
It should be clear, from this expression that even under  simplifying assumptions depicted in  Fig. (\ref{fig:Positions0}), the Hamiltonian constraint, a nonlinear elliptic PDE, cannot be solved exactly. Hence, we will resort to perturbation theory, but before  embarking on that computation, we can compute the ADM \cite{ADM} linear momentum and spin, using the exact extrinsic curvature, without any approximation. But of course we cannot compute the ADM energy.   This is because, assuming asymptotic flatness,  for the conformal factor, one has 
\begin{equation}
\psi(r)=1+\frac{E}{2r}+{\mathcal{O}}(1/r^{2})\hskip1cm{\text{as}}\,\,\,r\to\infty,
\label{asymp1}
\end{equation}
and defining the deviation  from the flat space as $h_{ij} :=(\psi^{4}-1)\delta_{ij}$, the total momentum of the hypersurface $\Sigma$ is determined only by the re-scaled extrinsic curvature on a sphere at infinity:  
\begin{equation}
P_{i}=\frac{1}{8\pi}\int_{S_{\infty}^{2}}dS\,n^{j}\,K_{ij}=\frac{1}{8\pi}\int_{S_{\infty}^{2}}dS\,n^{j}\,\hat{K}_{ij},\label{mom}
\end{equation}
which, for  (\ref{Bowen-York_extrinsiccurvature}) yields  $P_i = p_{1i}+ p_{2i}$. The total conserved {\it total angular momentum} is similar: 
\begin{equation}
J_{i}=\frac{1}{8\pi}\varepsilon_{ijk}\int_{S_{\infty}^{2}}dS\,n_{l}\,x^{j}K^{kl}=\frac{1}{8\pi}\varepsilon_{ijk}\int_{S_{\infty}^{2}}dS\,n_{l}\,x^{j}\hat{K}^{kl},\label{dad}
\end{equation}
yielding $J_{i}= j_{1i}+ j_{2i}$. On the other hand, to compute the ADM mass, we will need the 
exact form of the ${\mathcal{O}}(1/r)$ term in the conformal factor since we have 
\begin{equation}
E_{ADM}=\frac{1}{16\pi}\int_{S_{\infty}^{2}}dS\,n_{i}\,\Big(\partial_{j}h^{ij}-\partial_{i}h_{j}^{j}\Big)=-\frac{1}{2\pi}\int_{S_{\infty}^{2}}dS\,n^{i}\,\partial_{i}\psi,
\end{equation}
which we shall compute once we find the perturbative solution.

\section{Approximate solution of the Hamiltonian constraint for a binary black hole}

In order to solve (\ref{elliptic0}) with (\ref{BYkare}) on the right-hand side, we expand (\ref{BYkare}) up to and including ${\mathcal{O}}(p_i^2, j_i^2,c_i/r)$ which amounts to a slow moving, slow rotating binary and we are looking at regions away from the system as $c_1+c_2$ is the separation of black holes. After a slightly lengthy computation, the Hamiltonian constraint (\ref{elliptic0}) at this order becomes
\begin{eqnarray}
\hat{D}_{i}\hat{D}^{i}\psi &=& \psi^{-7}\Bigg[\frac{9\,p_{1}^{2}}{2\,r^{4}}\left( 1+2\sin^{2}{\theta} \,\sin^{2}\phi + \frac{12\,c_{1}}{r} \sin^{3}\theta \,\sin^{2}\phi  \, \cos\phi+\frac{4c_{1}}{r} \sin{\theta} \cos{\phi}  \right) \nonumber\\
&+& \frac{9\,p_{2}^{2}}{2\,r^{4}}\left( 1+2\sin^{2}\theta \,\sin^{2}\phi - \frac{12\,c_{2}}{r} \sin^{3}\theta \,\sin^{2}\phi  \, \cos\phi-\frac{4c_{2}}{r} \sin\theta \cos\phi  \right) \nonumber\\
&-& \frac{9\,p_{1} p_{2}}{r^{4}}\left( 1+2\sin^{2}\theta \,\sin^{2}\phi - \frac{6\ (c_{1}-c_{2})}{r} \sin^{3}\theta \,\sin^{2}\phi  \, \cos\phi + \frac{2(c_{1}-c_{2})}{r} \sin\theta \cos\phi  \right)\nonumber\\
&+&\frac{18\,j_{1}^{2}}{r^{6}}\left(\sin^{2}\phi + \cos^{2}\theta \cos^{2}\phi +\frac{8c_{1}}{r} \left(\sin\theta \sin^{2}\phi \cos \phi + \sin \theta \cos^{2}\theta \cos^{3}\phi    \right)\right)\nonumber\\ 
&+&\frac{18\,j_{2}^{2}}{r^{6}}\left(\sin^{2}\phi + \cos^{2}\theta \cos^{2}\phi -\frac{8c_{2}}{r} \left(\sin\theta \sin^{2}\phi \cos \phi + \sin \theta \cos^{2}\theta \cos^{3}\phi    \right)\right) \nonumber\\
&-&\frac{36\,j_{1}j_{2}}{r^{6}}\left(\sin^{2}\phi + \cos^{2}\theta \cos^{2}\phi +\frac{4(c_{1}-c_{2})}{r} \left(\sin\theta \sin^{2}\phi \cos \phi + \sin \theta \cos^{2}\theta \cos^{3}\phi    \right)\right) \nonumber\\
&+&\frac{18p_{1}j_{1}}{r^{5}}\left(\cos \theta+\frac{6c_{1}}{r}\sin {\theta} \cos {\theta} \cos{ \phi}\right) \nonumber\\
&+&\frac{18p_{2}j_{2}}{r^{5}}\left(\cos \theta-\frac{6c_{2}}{r}\sin {\theta} \cos {\theta} \cos{ \phi}\right) \nonumber\\
&-&\frac{18p_{1}j_{2}}{r^{5}}\left(\cos \theta+\frac{2(c_{1}-2c_{2})}{r}\sin {\theta} \cos {\theta} \cos{ \phi}\right) \nonumber\\
&-&\frac{18p_{2}j_{1}}{r^{5}}\left(\cos \theta+\frac{2(2c_{1}-c_{2})}{r}\sin {\theta} \cos {\theta} \cos{ \phi}\right)\Bigg].
\label{hamiltoian constraint}
\end{eqnarray}
Let us note that various sub-cases of this equation was studied before in the literature. An approximate solution of the Hamiltonian constraint for a {\it single}boosted slowly rotating gravitating system was given in \cite{Altas_Tekin_single}; and was elaborated in more detail in \cite{Altas_Tekin_review}. In \cite{Gleiser} a single slowly spinning black hole without linear momentum was solved in the leading order; and in \cite{Dennison-Baumgarte} a slowly moving  black hole without spin was studied. So the case we study here generalizes these previous works.

To be able to solve (\ref{hamiltoian constraint}) even in perturbation theory, one needs to make 
some judicious choices, otherwise the partial differential equations do not decouple. 
The form of the right-hand side of (\ref{hamiltoian constraint}) suggests a solution of the form 
\begin{eqnarray}
\psi(r,\theta,\phi)&:=&\psi^{(0)}+p_{1}^{2}\,\psi^{p_{1}^{2}}+p_{2}^{2}\,\psi^{p_{2}^{2}}+p_{1}p_{2}\, \psi^{p_{1}p_{2}}\nonumber\\
&+& j_{1}^{2}\,\psi^{j_{1}^{2}}+j_{2}^{2}\,\psi^{j_{2}^{2}}+j_{1}j_{2}\, \psi^{j_{1}j_{2}}\nonumber\\
&+&p_{1}j_{1}\, \psi^{p_{1}j_{1}}+p_{2}j_{2}\, \psi^{p_{2}j_{2}}+p_{1}j_{2}\, \psi^{p_{1}j_{2}}+p_{2}j_{1}\, \psi^{p_{2}j_{1}}+\ldots,
\label{expansionof psi}
\end{eqnarray}
where all the functions on the right-hand side depend on all the coordinates  $(r,\theta,\phi)$.
At the zeroth order, the right-hand side vanishes,  and the equation to be solved is the usual flat space Laplace equation 
\begin{equation}
\hat{D}_{i}\hat{D}^{i}\psi^{(0)}=0,
\end{equation}
which together with the
boundary conditions  \cite{Dennison-Baumgarte} at spatial infinity on $\Sigma$
\begin{equation}
\lim_{r\rightarrow\infty}\psi(r)=1,~~~~~~~~~\psi(r)>0,\label{boundarycondition1}
\end{equation}
and near the origin, has a unique solution
\begin{equation}
\lim_{r\rightarrow\ 0}\psi(r)=\psi^{(0)},\label{boundarycondition2}
\end{equation}
where $\psi^{(0)}$ might have a singularity at the origin.  In fact, the 
zeroth order solution satisfying these boundary conditions reads
\begin{equation}
\psi^{(0)}=1+\frac{a}{r}.
\end{equation}
Here $a$ is an integration constant which will appear in the ADM energy as is clear, but there will be additional contributions to the ADM energy coming form the spin and the linear momentum. The constant $a$ will also appear as the dominant term in the location of the apparent horizon. So its physical meaning will become transparent in the next section.
 On the right-hand side for the next order, one has
\begin{equation}
\psi^{-7} \sim (\psi^{(0)}) ^{-7}=\frac{r^{7}}{(r+a)^{7}},
\end{equation}
yielding the equations
\begin{eqnarray}
\hat{D}_{i}\hat{D}^{i}\psi^{p_{1}^{2}}=-\frac{9r^{3}}{16(r+a)^{7}}\left(1+2\sin^{2}\theta\, \sin^{2}\phi\right)-\frac{9r^{2}\,c_{1}}{4(r+a)^{7}}
\sin\theta\cos\phi\left(1+3\sin^{2}\theta\, \sin^{2}\phi\right),\label{p1p1equation}
 \end{eqnarray}
\begin{eqnarray}
\hat{D}_{i}\hat{D}^{i}\psi^{p_{2}^{2}}=-\frac{9r^{3}}{16(r+a)^{7}}\left(1+2\sin^{2}\theta\, \sin^{2}\phi\right)+\frac{9r^{2}\,c_{2}}{4(r+a)^{7}}
\sin\theta\cos\phi\left(1+3\sin^{2}\theta\, \sin^{2}\phi\right),\label{p2p2equation}
 \end{eqnarray}
\begin{eqnarray}
\hat{D}_{i}\hat{D}^{i}\psi^{p_{1}p_{2}}=\frac{9r^{3}}{8(r+a)^{7}}\left(1+2\sin^{2}\theta\, \sin^{2}\phi\right)+\frac{9r^{2}(c_{1}-c_{2})}{4(r+a)^{7}}
\sin\theta\cos\phi\left(1+3\sin^{2}\theta\, \sin^{2}\phi\right),\label{p1p2equation}
 \end{eqnarray}
\begin{eqnarray}
\hat{D}_{i}\hat{D}^{i}\psi^{j_{1}^{2}}=\frac{-9r}{4(r+a)^{7}}(\sin^{2}\phi+\cos^{2}\theta\cos^{2}\phi)-\frac{18 c_{1}}{(r+a)^{7}}\sin\theta\cos^3\phi(\tan^{2}\phi+\cos^{2}\theta),
\label{j1j1equation}
 \end{eqnarray}
\begin{eqnarray}
\hat{D}_{i}\hat{D}^{i}\psi^{j_{2}^{2}}=\frac{-9r}{4(r+a)^{7}}(\sin^{2}\phi+\cos^{2}\theta\cos^{2}\phi)+\frac{18 c_{2}}{(r+a)^{7}}\sin\theta\cos^3\phi(\tan^{2}\phi+\cos^{2}\theta),
\label{j2j2equation}
 \end{eqnarray}
\begin{eqnarray}
\hat{D}_{i}\hat{D}^{i}\psi^{j_{1}j_{2}}=&&\frac{9r}{2(r+a)^{7}}(\sin^{2}\phi+\cos^{2}\theta\cos^{2}\phi) 
+\frac{18 (c_{1}-c_{2})}{(r+a)^{7}}\sin\theta\cos^3\phi(\tan^{2}\phi+\cos^{2}\theta),
\label{j1j2equation}
 \end{eqnarray}
\begin{eqnarray}
\hat{D}_{i}\hat{D}^{i}\psi^{p_{1}j_{1}}=\frac{-9r^{2}}{4(r+a)^{7}}\cos\theta -\frac{27r c_{1}}{2(r+a)^{7}}\sin\theta\cos\theta\cos\phi,
\label{p1j1equation}
 \end{eqnarray}
\begin{eqnarray}
\hat{D}_{i}\hat{D}^{i}\psi^{p_{2}j_{2}}=\frac{-9r^{2}}{4(r+a)^{7}}\cos\theta +\frac{27r c_{2}}{2(r+a)^{7}}\sin\theta\cos\theta\cos\phi,
\label{p2j2equation}
 \end{eqnarray}
\begin{eqnarray}
\hat{D}_{i}\hat{D}^{i}\psi^{p_{1}j_{2}}=\frac{9r^{2}}{4(r+a)^{7}}\cos\theta +\frac{9r (c_{1}-2c_{2})}{2(r+a)^{7}}\sin\theta\cos\theta\cos\phi,
\label{p1j2equation}
 \end{eqnarray}
\begin{eqnarray}
\hat{D}_{i}\hat{D}^{i}\psi^{p_{2}j_{1}}=\frac{9r^{2}}{4(r+a)^{7}}\cos\theta +\frac{9r (2c_{1}-c_{2})}{2(r+a)^{7}}\sin\theta\cos\theta\cos\phi.
\label{p2j1equation}
 \end{eqnarray}
Each equation, albeit being linear, is still a PDE; one can convert these equations to decoupled ODEs with the help of the
following spherical harmonics:
\begin{eqnarray*}
 &  & Y_{0}^{0}(\theta,\phi)=\frac{1}{\sqrt{4\pi}},~~~~~~~~~~~~~~~Y_{1}^{0}(\theta,\phi)=\sqrt{\frac{3}{4\pi}}\cos\theta,~~~~~~Y_{2}^{0}(\theta,\phi)=\sqrt{\frac{5}{16\pi}}(3\cos^{2}\theta-1),\\
 &  & Y_{1}^{-1}(\theta,\phi)=\sqrt{\frac{3}{4\pi}}\sin\theta\sin\phi,~~Y_{2}^{1}(\theta,\phi)=\sqrt{\frac{15}{4\pi}}\sin\theta\cos\theta\cos\phi,~~Y_{1}^{1}(\theta,\phi)=\sqrt{\frac{3}{4\pi}}\sin\theta\cos\phi.
\end{eqnarray*}
The ansatz for (\ref{p1p1equation}) is of the form:
\begin{eqnarray}
\psi^{p_{1}^{2}}(r,\theta,\phi)&=&\psi_{0}^{p_{1}^{2}}(r)\big[Y_{0}^{0}(\theta,\phi)\big]^{2}+\psi_{1}^{p_{1}^{2}}(r)\big[Y_{1}^{-1}(\theta,\phi)\big]^{2}\nonumber\\
&&+c_{1}\psi_{2}^{p_{1}^{2}}(r)\big[Y_{1}^{1}(\theta,\phi)\big]^{2}+c_{1}\psi_{3}^{p_{1}^{2}}(r)Y_{1}^{1}(\theta,\phi)\big[Y_{1}^{-1}(\theta,\phi)\big]^{2}.
\label{ansatzp1p1}
\end{eqnarray}
As this structure shows, the spherical harmonics enter into the picture in a rather non-trivial way, one has to make careful choices to decouple the radial and angular parts. We do not depict here 
the solutions to the radial parts separately, as the expressions become rather long. The solution to 
(\ref{p1p1equation}), obeying the boundary conditions, for the ansatz (\ref{ansatzp1p1}), turns out to be 

\begin{eqnarray}
&&\psi^{p_{1}^{2}}(r,\theta,\phi)=-\frac{84 a^6+378 a^5 r+653 a^4 r^2+514 a^3 r^3+142 a^2 r^4-35 a r^5-25 r^6}{160 a r^2 (a+r)^5}\nonumber \\
&& +\frac{21 a \log \left(\frac{a+r}{a}\right)}{40 r^3} -\frac{c_1 \sin \theta \cos \phi } {80 a r^4 (a+r)^5}\times \nonumber \\
&&\left(108 a^2 (a+r)^5 \log \left(\frac{a}{a+r}\right)+r \left(108 a^6+486 a^5 r+846 a^4 r^2+693 a^3 r^3+245 a^2 r^4+10 a r^5-16 r^6\right)\right)\nonumber \\
&&+\frac{\sin ^2 \theta  \sin ^2 \phi }{160 r^3} \times \nonumber \\
&& 3 \left(\frac{r \left(84 a^5+378 a^4 r+658 a^3 r^2+539 a^2 r^3+192 a r^4+15 r^5\right)}{(a+r)^5}+84 a \log \left(\frac{a}{a+r}\right)\right) \nonumber \\
&&+\frac{9 c_1 \sin ^3 \theta  \sin ^2 \phi  \cos \phi }{80 r^4 (a+r)^5} \times  \\
&& \Bigg(r \left(60 a^5+270 a^4 r+470 a^3 r^2+385 a^2 r^3+137 a r^4+10 r^5\right)+60 a (a+r)^5 \log \left(\frac{a}{a+r}\right) \Bigg). \nonumber 
\end{eqnarray}
$\psi^{p_{2}^{2}}(r,\theta,\phi)$ can be obtained from the above expression via the replacement $ c_1 \rightarrow -c_2$. So we do not depict it here. The solution to  (\ref{p1p2equation}) is 
\begin{eqnarray}
&&\psi^{p_{1}p_{2}}(r,\theta,\phi)=\frac{84 a^6+378 a^5 r+653 a^4 r^2+514 a^3 r^3+142 a^2 r^4-35 a r^5-25 r^6}{80 a r^2 (a+r)^5}\nonumber \\
&& -\frac{21 a \log \left(\frac{a+r}{a}\right)}{20 r^3} +\frac{(c_1 -c_2) \sin \theta \cos \phi } {80 a r^4 (a+r)^5}\times \nonumber \\
&&\left(108 a^2 (a+r)^5 \log \left(\frac{a}{a+r}\right)+r \left(108 a^6+486 a^5 r+846 a^4 r^2+693 a^3 r^3+245 a^2 r^4+10 a r^5-16 r^6\right)\right)\nonumber \\
&&-\frac{\sin ^2 \theta  \sin ^2 \phi }{80 r^3} \times \nonumber \\
&& 3 \left(\frac{r \left(84 a^5+378 a^4 r+658 a^3 r^2+539 a^2 r^3+192 a r^4+15 r^5\right)}{(a+r)^5}+84 a \log \left(\frac{a}{a+r}\right)\right) \nonumber \\
&&-\frac{9 (c_1-c_2) \sin ^3 \theta  \sin ^2 \phi  \cos \phi }{160r^4 (a+r)^5} \times \\
&& \Bigg(r \left(60 a^5+270 a^4 r+470 a^3 r^2+385 a^2 r^3+137 a r^4+10 r^5\right)+60 a (a+r)^5 \log \left(\frac{a}{a+r}\right) \Bigg).\nonumber 
\end{eqnarray}
The ansatz for (\ref{j1j1equation}) is of the form:
\begin{eqnarray}
\psi^{j_{1}^{2}}(r,\theta,\phi)&=&\psi_{0}^{j_{1}^{2}}(r)\big[Y_{0}^{0}(\theta,\phi)\big]^{2}+\psi_{1}^{j_{1}^{2}}(r)\big[Y_{1}^{1}(\theta,\phi)\big]^{2}\nonumber\\
&&+c_{1}\psi_{2}^{j_{1}^{2}}(r)\big[Y_{1}^{1}(\theta,\phi)\big]+c_{1}\psi_{3}^{j_{1}^{2}}(r)\big[Y_{1}^{1}(\theta,\phi)\big]^{3},
\label{ansatzj1j1}
\end{eqnarray}
and the corresponding solution is;
\begin{eqnarray}
&&\psi^{j_{1}^{2}}(r,\theta,\phi)=-\frac{a^{4}+5a^{3}r+11a^{2}r^{2}+5ar^{3}+r^{4}}{40a^{3}(a+r)^{5}}+\frac{\sin\theta\cos\phi\,r\,c_{1}}{10a^{8}(a+r)^{5}}\times \nonumber\\
&&\left(a(4a^{6}+20a^{5}r+247a^{4}r^{2}+693a^{3}r^{3}+846a^{2}r^{4}+486ar^{5}+108r^{6})+108r^{2}(a+r)^{5}(\log\frac{r}{a+r})\right)\nonumber\\
&&-\frac{\sin^{2}\theta\cos^{2}\phi\,3r^{2}}{40a(a+r)^{5}}-\frac{3\sin^{3}\theta\cos^{3}\phi\, c_{1}r^{2}}{10a^{8}(a+r)^{5}}\times \\
&&\left(a(10a^{5}+137a^{4}r+385a^{3}r^{2}+470a^{2}r^{3}+270ar^{4}+60r^{5})+60r(a+r)^{5}\log\frac{r}{a+r} \right).\nonumber 
\end{eqnarray}
Ansatzes for (\ref{j2j2equation}) and (\ref{j1j2equation}) are the same as (\ref{j1j1equation}) after  the substitutions $c_{1}\rightarrow -c_{2}$ and $c_{1}\rightarrow (c_{2}-c_{1})$, respectively. Hence we do not depict them here.

The ansatz for (\ref{p1j1equation}) is of the form:
\begin{eqnarray}
\psi^{p_{1}j_{1}}(r,\theta,\phi)&=&\psi_{0}^{p_{1}j_{1}}(r)\big[Y_{0}^{0}(\theta,\phi)]^{2}+\psi_{1}^{p_{1}j_{1}}(r)Y_{1}^{0}(\theta,\phi)+c_{1}\psi_{2}^{p_{1}j_{1}}(r)Y_{2}^{1}(\theta,\phi),
\label{ansatzp1j1}
\end{eqnarray}
and the corresponding solution is;
\begin{equation}
\psi^{p_{1}j_{1}}(r,\theta,\phi)=\frac{r(a^{2}+5ar+10r^{2})\cos\theta}{80a(a+r)^{5}}+\frac{9c_{1}r^{2}\sin\theta\cos\theta\cos\phi}{20a(a+r)^{5}}.
\end{equation}
The ansatzes for (\ref{p2j2equation}),  (\ref{p1j2equation}) and (\ref{p2j1equation}) are the same as  (\ref{ansatzp1j1}) after the change of coefficients $c_{1}\rightarrow -c_{2}$,  $c_{1}\rightarrow (2c_{2}-c_{1})/3$ and  $c_{1}\rightarrow (c_{2}-2c_{3})/3$ respectively, and the corresponding solutions are:
\begin{equation}
\psi^{p_{2}j_{2}}(r,\theta,\phi)=\frac{r(a^{2}+5ar+10r^{2})\cos\theta}{80a(a+r)^{5}}-\frac{9c_{2}r^{2}\sin\theta\cos\theta\cos\phi}{20a(a+r)^{5}},
\end{equation}
\begin{equation}
\psi^{p_{1}j_{2}}(r,\theta,\phi)=\frac{-r(a^{2}+5ar+10r^{2})\cos\theta}{80a(a+r)^{5}}-\frac{3(c_{1}-2c_{2})r^{2}\sin\theta\cos\theta\cos\phi}{20a(a+r)^{5}},
\end{equation}
\begin{equation}
\psi^{p_{2}j_{1}}(r,\theta,\phi)=\frac{-r(a^{2}+5ar+10r^{2})\cos\theta}{80a(a+r)^{5}}-\frac{3(2c_{1}-c_{2})r^{2}\sin\theta\cos\theta\cos\phi}{20a(a+r)^{5}}.
\end{equation}

Collecting all the pieces above and inserting them into (\ref{expansionof psi}), one finds the conformal factor from which all the ensuing computations will follow. But the explicit form is unwieldy. We have used Mathematica to keep track of the computations and check all the details. Let us show the expansion of the conformal factor as a solution to the Hamiltonian constraint up to ${\mathcal{O}}(1/r^3)$:
\begin{eqnarray}
\psi(r,\theta,\phi)&&=1+\frac{160 a^4+25 a^2 {\mathcal {P}}_r^2+4 {\mathcal{J}}_r^2}{160 a^3 r} + \frac{{\mathcal {P}}_r \Big (16{\mathcal{J}}_r \cos \theta -9 a {\mathcal {P}}_r (\cos 2 \theta +7)\Big)}{128 a r^2} \nonumber \\
&&+\frac{\sin \theta \Big(64 \cos \phi  \left(5 a^2 {\mathcal{P}} {\mathcal {P}}_r+{\mathcal{J}} {\mathcal{J}}_r\right)-225 a^3 {\mathcal {P}}_r^2 \sin \theta  \cos 2 \phi \Big)}{1600 a^3 r^2}+ {\mathcal{O}}(1/r^3), \label{loworder}
\end{eqnarray}
where we have defined
\begin{equation}
{\mathcal {P}}_r := p_1- p_2, \hskip 1 cm  {\mathcal{P}}:= p_1 c_1 + p_2 c_2, \hskip 1 cm   {\mathcal{J}}_r := j_1- j_2 \hskip 1 cm {\mathcal{J}}:= j_1 c_1 + j_2 c_2.
\label{denkb}
\end{equation}
To compute the ADM energy, we need the $\mathcal{O}(\frac{1}{r})$ in (\ref{loworder}), which from
(\ref{asymp1})  yields
\begin{equation}
\boxed{E_{\text{ADM}}=2a+\frac{5 {\mathcal {P}}_r^{2}}{16a}+\frac{{\mathcal{J}}_r ^{2}}{20a^{3}}.}
\end{equation}
For vanishing spin and vanishing linear
momentum, it is clear that the constant $a$ is related to the  total mass of the static spacetime.
In the next section, we will write the ADM mass in terms of the irreducible mass once we find the apparent horizon.

\section{Finding the Apparent Horizon}

\begin{figure}
\centering \includegraphics[width=0.7\linewidth]{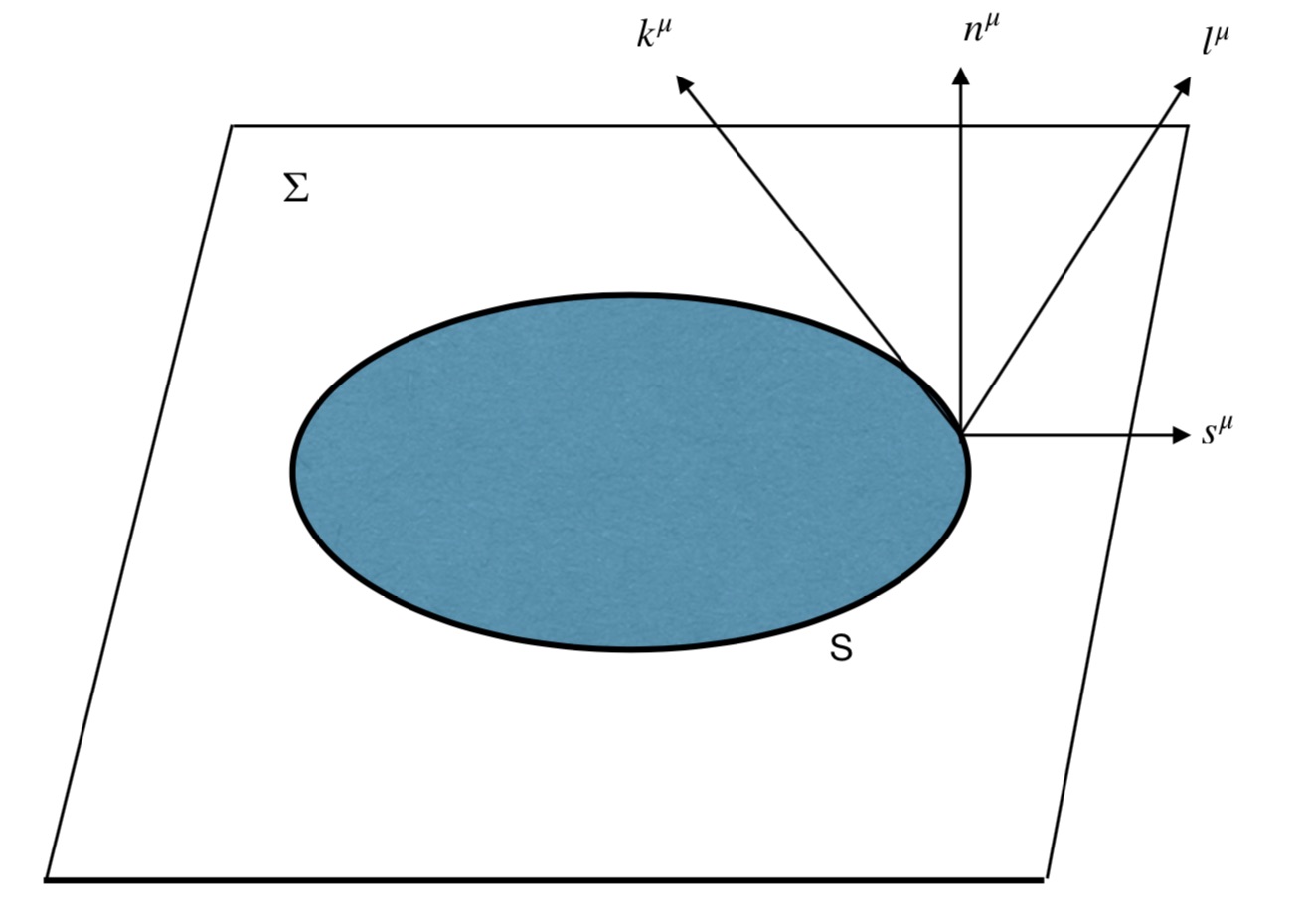}\caption{A schematic picture of the apparent horizon, the boundary of the colored region. The colored region is a trapped region and the apparent horizon is a marginally trapped surface. $n^\mu$ is timelike, $s^\mu$ is spacelike, while $k^\mu$ and $\ell^\mu$ are lightlike. Lie-dragging the metric on the apparent horizon along the outgoing null vector $\ell$ yields zero; and also derivative of the area of the apparent horizon along $\ell$ gives zero. }
\label{fig:Positions} 
\end{figure}

The apparent horizon is a codimension two spacelike hypersurface (unlike the event horizon which is a codimension one null hypersurface). Since it is discussed at length in the literature  (see  \cite{Baumgarte} and \cite{Altas_Tekin_review}), we shall only briefly recap the relevant equations. Assume now that  $(r,\theta,\phi) $ are local coordinates on the apparent horizon 
${\mathcal{S}}$, a subspace of $\Sigma$ with the pull-back metric $q_{\mu \nu}$ from the spacetime metric.  And $s^i$ is a unit normal to the surface, as shown in Fig. (\ref{fig:Positions}). Then the apparent horizon equation reads 
\begin{equation}
q^{ij}\left(\partial_{i}s_{j}-^{\Sigma}\Gamma_{ij}^{k}s_{k}-K_{ij}\right)=0.\label{equationinaxisymmetry0}
\end{equation}
Furthermore, assume that the surface ${\mathcal{S}}$ can be parameterized as a
level set of a function $\Phi(r,\theta,\phi)$ such that 
\begin{equation}
\Phi(r,\theta,\phi):=r-h(\theta,\phi)=0,
\end{equation}
with $h$ being a smooth function. A rather tedious computation yields the following exact equation for a conformally flat, maximally sliced hypersurface:
\begin{eqnarray}
 &  & -\gamma^{\theta\theta}\partial_{\theta}^{2}h-\gamma^{\phi\phi}\partial_{\phi}^{2}h-\frac{1}{2}\Bigl((\gamma^{rr})^{2}\partial_{r}\gamma_{rr}-\gamma^{\theta\theta}\gamma^{rr}\partial_{r}\gamma_{\theta\theta}-\gamma^{\phi\phi}\gamma^{rr}\partial_{r}\gamma_{\phi\phi}+\partial_{\theta}h\gamma^{\phi\phi}\gamma^{\theta\theta}\partial_{\theta}\gamma_{\phi\phi}\Bigr)\nonumber \\
 &  & +\lambda^{2}\Bigl((\gamma^{\theta\theta})^{2}(\partial_{\theta}h)^{2}\partial_{\theta}^{2}h+(\gamma^{\phi\phi})^{2}(\partial_{\phi}h)^{2}\partial_{\phi}^{2}h+2\gamma^{\phi\phi}\gamma^{\theta\theta}\partial_{\phi}h\partial_{\theta}h\partial_{\theta}\partial_{\phi}h\Bigr)\nonumber \\
 &  & +\frac{\lambda^{2}}{2}\Bigl((\gamma^{rr})^{3}\partial_{r}\gamma_{rr}+(\gamma^{\theta\theta})^{2}\gamma^{rr}(\partial_{\theta}h)^{2}\partial_{r}\gamma_{\theta\theta}+(\gamma^{\phi\phi})^{2}\gamma^{rr}(\partial_{\phi}h)^{2}\partial_{r}\gamma_{\phi\phi}\nonumber \\
 &  & ~~~~~~~~~~~~~-(\partial_{\phi}h)^{2}\partial_{\theta}h(\gamma^{\phi\phi})^{2}\gamma^{\theta\theta}\partial_{\theta}\gamma_{\phi\phi}\Bigr)\nonumber \\
 &  & +\lambda\Bigl((\gamma^{rr})^{2}K_{rr}+(\gamma^{\theta\theta})^{2}(\partial_{\theta}h)^{2}K_{\theta\theta}+(\gamma^{\phi\phi})^{2}(\partial_{\phi}h)^{2}K_{\phi\phi}-2\gamma^{rr}\gamma^{\theta\theta}\partial_{\theta}hK_{r\theta}\nonumber \\
 &  & ~~~~~~~~~~~~~~-2\gamma^{rr}\gamma^{\phi\phi}\partial_{\phi}hK_{r\phi}+2\gamma^{\theta\theta}\gamma^{\phi\phi}\partial_{\theta}h\partial_{\phi}hK_{\theta\phi}\Bigr)=0,\label{yarab}
\end{eqnarray}
where $\lambda$ is given as
\begin{equation}
\lambda=\Big(\gamma^{rr}+\gamma^{\theta\theta}(\partial_{\theta}h)^{2}+\gamma^{\phi\phi}(\partial_{\phi}h)^{2}\Big)^{-1/2}.
\end{equation}
In principle, given the initial data, {\it i.e.} $\gamma_{ij}$ and $K_{ij}$, one can solve (\ref{yarab}) numerically.  But we shall attempt a perturbative solution, consistent with our approach so far.  For this purpose, we need the extrinsic curvature in the spherical coordinates. So we make a coordinate transformation from the Cartesian coordinates to the spherical coordinates. Then $\hat{K}_{i j}$ transforms in to a spherical tensor of which the non-vanishing components are
\begin{eqnarray}
\hat{K}_{rr}&=&\frac{3}{r^{2}}{\mathcal{P}}_r \sin\theta\sin\phi+\frac{6}{r^{3}}{\mathcal{P}}\sin^{2}\theta\sin\phi\cos\phi,\nonumber\\
\hat{K}_{r\theta}&=&\frac{3}{2r}{\mathcal{P}}_r\cos\theta\sin\phi+\frac{3}{r^{2}}{\mathcal{J}}_r\sin\phi+\frac{12}{r^{3}}{\mathcal{J}}\sin\theta\sin\phi\cos\phi, \nonumber\\
\hat{K}_{r\phi}&=&\frac{3}{r^{2}}{\mathcal{P}}_r\sin\theta\cos\phi+\frac{3}{r^{2}}{\mathcal{P}}\sin^{2}\theta 
+ \frac{3}{r^{2}}{\mathcal{J}}_r\sin\theta\cos\theta\cos\phi+ \frac{12}{r^{3}}{\mathcal{J}}\sin^{2}\theta\cos\theta\cos^{2}\phi,
\label{radyalextrinsic}
\end{eqnarray}
where $\mathcal{P}$ {\it etc.} were defined in (\ref{denkb}).

Following Christodoulou \cite{Chris} the irreducible mass $M_{\text{irr}}$ of the black hole can be defined in terms of the area of a cross-section of the event horizon as 
\begin{equation}
M_{\text{irr}}:=\sqrt{\frac{A_{\text{EH}}}{16\pi}}.\label{irreduciblemassformula0}
\end{equation}
For the non-stationary case that we are dealing with, instead of a section of the event horizon, we can use the  apparent horizon \cite{Dennison-Baumgarte} as a viable approximation, hence we have 
\begin{equation}
M_{\text{irr}}:=\sqrt{\frac{A_{\text{AH}}}{16\pi}},\label{irreduciblemassformula}
\end{equation}
where the exact area reads
\begin{equation}
A_{\text{AH}}=\intop_{0}^{2\pi}d\phi\intop_{0}^{\pi}d\theta\thinspace\sin\theta\thinspace\psi^{4}\thinspace (h+a)^{2}\left(1+\frac{1}{(h+a)^{2}}\left(\partial_{\theta}h\right)^{2}+\frac{1}{(h+a)^{2}\sin^{2}\theta}\left(\partial_{\phi}h\right)^{2}\right)^{1/2}.\label{alan3}
\end{equation}
To compute this at the order we are working, we need to solve (\ref{yarab}) up to first order in the parameters $p_{1},p_{2},j_{1},j_{2}$; therefore plugging the  ansatz 
\begin{equation}
h(\theta,\phi)=h^{0}+p_{1}h^{p_{1}}++p_{2}h^{p_{2}}+j_{1}h^{j_{1}}+j_{2}h^{j_{2}}+\mathcal{O}(p_{1}^{2},p_{2}^{2},j_{1}^{2},j_{2}^{2},\ldots)
\end{equation}
into (\ref{yarab}), one arrives at 
\begin{eqnarray}
\frac{-\psi^{4}}{r^{2}}\left(\partial_{\theta}^{2}h+\frac{\partial_{\phi}^{2}h}{\sin^{2}\theta}+\cot\theta\partial_{\theta}h-2r-4r^{2}\frac{\partial_{r}\psi}{\psi}\right)
\hat{K}_{rr}-2r^{-2}\partial_{\theta}h \hat{K}_{r\theta}-2r^{-2}\sin^{2}\theta \partial_{\phi}h \hat{K}_{r\phi}=0.
\label{result1}
\end{eqnarray}
Substituting (\ref{radyalextrinsic}) and the conformal factor derived in the previous section,  into (\ref{result1}) gives the following differential equations;
\begin{eqnarray}
&&\partial^{2}_{\theta}h^{p_{1}}+\frac{1}{\sin^{2}\theta} \partial^{2}_{\phi}h^{p_{1}}+\cot\theta \partial_{\theta}h^{p_{1}}-h^{p_{1}}-\frac{3}{16}\sin\theta\sin\phi -\frac{3}{8a}c_{1}\sin^{2}\theta\sin\phi\cos\phi=0, \nonumber\\
&&\partial^{2}_{\theta}h^{p_{2}}+\frac{1}{\sin^{2}\theta} \partial^{2}_{\phi}h^{p_{2}}+\cot\theta \partial_{\theta}h^{p_{2}}-h^{p_{2}}+\frac{3}{16}\sin\theta\sin\phi -\frac{3}{8a}c_{2}\sin^{2}\theta\sin\phi\cos\phi=0, \nonumber\\
&&\partial^{2}_{\theta}h^{j_{1}}+\frac{1}{\sin^{2}\theta} \partial^{2}_{\phi}h^{j_{1}}+\cot\theta \partial_{\theta}h^{j_{1}}-h^{j_{1}}=0, 
\label{apparentdiffeq}
\end{eqnarray}
with $j_2$ satisfying the same equation as the last one. 
At the zeroth order, $\mathcal{O}(p^{0},J^{0})$, one has the solution
\begin{equation}
h^{0}=a,
\end{equation}
which shows that $a$ as the location of
the apparent horizon at the lowest order. The remaining equations are of the  homogeneous and non-homogeneous  Helmholtz equations on the two sphere ($S^{2}$): 
\begin{equation}
\left({\vec{\nabla}}_{S^{2}}^{2}+k\right)f\left(\theta,\phi\right)=g\left(\theta,\phi\right),
\end{equation}
with the Laplacian on the sphere given as 
\begin{equation}
{\vec{\nabla}}_{S^{2}}^{2}:=\partial_{\theta}^{2}+\cot\theta\partial_{\theta}+\frac{1}{\sin^{2}\theta}\partial_{\phi}^{2}.
\end{equation}
In \cite{Altas_Tekin_single} and \cite{Altas_Tekin_review}, we described how this equation can be solved via the Green's function technique; here we do not repeat that computation, instead just write the result: at this order the apparent horizon is
given by the solution
\begin{equation}
\boxed{h(\theta,\phi)= a+\frac{(p_1-p_2)}{16} \sin \theta \sin \phi -\frac{3(p_1 c_1+ p_2 c_2) \sin ^2 \theta \sin \phi \cos \phi }{56 a}.}
\label{AHdenk}
\end{equation}
Using (\ref{AHdenk}),  the area of the apparent horizon can be calculated  from (\ref{alan3}) to get
\begin{equation}
A_{\text{AH}}=64\pi a^{2}+4\pi \left(p_{1}-p_{2}\right)^{2}+\frac{11\pi}{5a^{2}}\left(j_{1}-j_{2}\right)^{2}.
\end{equation}
Therefore the irreducible mass from (\ref{irreduciblemassformula})  is 
\begin{equation}
M_{\text{irr}}=2a+\frac{(p_{1}-p_{2})^{2}}{16a}+\frac{11(j_{1}-j_{2})^{2}}{320a^{3}}.
\end{equation}
The $E_{ADM}$ energy can be expressed in terms of the irreducible mass as  
\begin{equation}
E_{\text{ADM}}=M_{\text{irr}}+\frac{(p_{1}-p_{2})^{2}}{2M_{\text{irr}}}+\frac{(j_{1}-j_{2})^{2}}{8M_{\text{irr}}^{3}},
\end{equation}
which matches the result of \cite{Chris} at this
order.

\section{Conclusions and Discussions}

Extending our earlier work \cite{Altas_Tekin_single}, in which we analytically, albeit perturbatively, found a single boosted, rotating gravitating system as an initial data for Einstein's theory in a vacuum; here we have studied binary black holes with total spin and linear momentum orbiting around each other. We worked in the Bowen-York formalism where the momentum constraints decouple and admit exact solutions, while the Hamiltonian constraint, a nonlinear elliptic equation, is solved perturbatively. We determined the conformal factor for small momenta and rotation and for close separation of black holes. We have also determined the shape of the apparent horizon covering both black holes in close separation, as well as the conserved quantities, such as energy, momentum, angular momentum and irreducible mass associated with the solution. For an earlier work on close-limit of binary black hole collisions and the associated radiation in the context of Misner initial data \cite{Misner}, see  \cite{Price}. Our work is valid for distances far away from the binary system, the analytical results presented here can be used to check the numerical computations for far distances. Numerical relativity results should match our analytical results in this regime for the particular black hole configurations described as in Fig. 1. Our solution can be criticized on the basis that numerical methods used for the evolution equations are far superior to describe the merging black holes even in the close proximity where our perturbative scheme is insufficient. While we agree with this, we worked with the constraint equations, and not the time evolution equations; and also
it is always good to have an analytical description of the system even if that description is valid in some perturbative regime such as we have here.   For further work, one can start with the exact Hamilton's constraint (\ref{elliptic0}) of which the right-hand side is given by (\ref{BYkare}) and try to either solve in some approximation or numerically for generic orientation and separation of black holes. We have not been able to do that so far, but any further improvement in that direction would be interesting.

\begin{acknowledgments}
The work of E.A. and E.E is partially supported by the TUBITAK Grant No. 120F253.  B.T. would like to dedicate this work to the 13 close family members, with ages between 3 and 56,  who died in the massive earthquakes that hit southeastern part of Turkey on 6 February 2023.
\end{acknowledgments}

\end{document}